# Performance of the Two Aerogel Cherenkov Detectors of the JLab Hall A Hadron Spectrometer

S. Marrone[*a], B. B. Wojtsekhowski[b], A. Acha[c], E. Cisbani[d], M. Coman[c], F. Cusanno[e], C. W. de Jager[b], R. De Leo[f], H. Gao[†g], F. Garibaldi[d], D. W. Higinbotham[b], M. Iodice[h], J. J. LeRose[e], D. Macchia[f], P. Markowitz[c], E. Nappi[a], F. Palmisano[i], G. M. Urciuoli[e], I. van der Werf[f], H. Xiang[g], L. Y. Zhu[g]

a) Istituto Nazionale di Fisica Nucleare Sezione di Bari, Bari, Italy.
b) Thomas Jefferson National Accelerator Facility, Newport News, Virginia, USA.
c) Florida International University, Miami, Florida, USA.
d) Istituto Nazionale di Fisica Nucleare, Sez. Roma, gruppo Sanità, and Istituto Superiore di Sanità, Rome, Italy.
e) Istituto Nazionale di Fisica Nucleare, Sezione di Roma, Rome, Italy.
f) Dipartimento di Fisica, Università di Bari, and Istituto Nazionale di Fisica Nucleare Bari, Italy.
g) Massachussets Institute of Technology, Cambridge, Massachussets, USA.
h) Istituto Nazionale di Fisica Nucleare, Sezione di Roma Tre, Rome, Italy.
i) Dipartimento di Chimica, Università di Bari, Bari, Italy.

Date: 23 October 2008.

**Abstract**

We report on the design and commissioning of two silica aerogel Cherenkov detectors with different refractive indices. In particular, extraordinary performance in terms of the number of detected photoelectrons was achieved through an appropriate choice of PMT type and reflector, along with some design considerations. After four years of operation, the number of detected photoelectrons was found to be noticeably reduced in both detectors as a result of contamination, yellowing, of the aerogel material. Along with the details of the set-up, we illustrate the characteristics of the detectors during different time periods and the probable causes of the contamination. In particular we show that the replacement of the contaminated aerogel and parts of the reflecting material has almost restored the initial performance of the detectors.

## 1. Introduction

To perform advanced high resolution hypernuclear spectroscopy [1] in Hall A at the Thomas Jefferson National Accelerator Facility (JLab), several challenging instruments have been developed. The low cross section (e,e'K) experiments [1-3] were performed using high-resolution spectrometers with an electron beam of high energy and high luminosity. In those experiments the coincidence events resulting from the (e,e'π) reaction exceed by several orders of magnitude the hypernuclear event signals, (e,e'K), therefore this type of measurement needs a powerful particle identification (PID) system. In this experiment a four-component PID system was implemented in Hall A, a gas Cherenkov detector [4] for the electron arm while in the hadron arm a high resolution

---

[*] Corresponding author: e-mail stefano.marrone@ba.infn.it, Fax. +390805442470, Tel. +390805442511.
[†] Current address: Department of Physics and the Triangle Universities Nuclear Laboratory, Duke University, Durham, NC 27708.

Time-of-Flight scintillator detector [5,6], threshold Cherenkov detectors with two different refraction indices [7], and a Ring Imaging Cherenkov (RICH) detector [8] are installed. Together they provide a rejection factor of more than 10,000 for protons and pions. In this paper we present detailed information about the threshold aerogel detectors and the related technique of photodetection, which is in many aspects a continuation of earlier developments, see for a review Ref. [9]. In particular it is shown that with an appropriate choice of the reflector and photomultipliers, and some innovative design considerations, extraordinary performance in the number of detected photoelectrons was achieved. This performance was necessary for the success of some experiments in Hall A [1-3] and more recently was exploited in the construction of the aerogel detectors installed in the High Momentum Spectrometer of the Hall C [10].

The upgrade (to a 12 GeV beam energy) of the Continuous Electron Beam Accelerator Facility (CEBAF) [11] will enhance the potential for new experiments at JLab. It will open unique opportunities for investigation of hadron structure by means of Semi-Inclusive Deep Inelastic Scattering (SIDIS) at high luminosity. High luminosity will also allow a more detailed study of the (e,e'K) reaction in SIDIS kinematics by putting the hadron-PID in the centre of the detector system. In this scenario threshold aerogel Cherenkov detectors are required to operate an on-line PID at the trigger level. A RICH provides better hadron identification for off-line analysis.

Hall A is equipped with two High Resolution Spectrometers (HRS); one to detect the scattered electron, the other for the produced hadron [12], where particle identification is provided by the previously mentioned PID system. HRS are able to select particle momenta up to 4.3 GeV/$c$ with a resolution better than $2\times10^{-4}$ and with a horizontal angular uncertainty of 2 mrad for a central momentum of 4 GeV/$c$. With such characteristics, Hall A is well suited for high luminosity, high energy resolution, exclusive and semi-exclusive electron scattering experiments [1-3].

In 2000 two Cherenkov detectors (A1 and A2), with radiators of silica aerogel having different refractive indices, were installed in the focal plane of the hadron HRS for the identification of pions, protons, and kaons in the E94-104 experiment [13]. In the 2001-2003 years, other Hall A experiments [2,3] successfully employed these detectors with performance qualitatively close to that of the original installation (2000). In 2004, during the first runs of the experiment E94-107 [1], the photoelectron (PE) distribution, measured by A1 and A2, was found to have a noticeably reduced average number of PE's. This degradation was due to the yellowing of the diffusive reflector, covering the detector internal surface, and to the outer layers of aerogel, which on visual inspection appeared less transparent. The sources of contamination were initially suspected to be the double sided tape, used to hold the Millipore reflector, and the air forced to flush through the detectors. In order to have the detectors performing optimally for the high precision hypernuclear experiment E94-107 in 2005, the degraded aerogel layers and the reflector material were replaced.

This paper reports on the construction of the detectors and on their performance during three periods: after the first installation (2000 original configuration), at the time of the reduced PE yield (2004 degraded configuration), and soon after the restoring procedures (2005 restored configuration).

Sect. 2 describes the detectors and provides the details of the design considerations, particularly the optical properties of the silica aerogel employed, the mechanical assembly, the reflecting materials, and the PMT's used. In particular the causes of the contamination are analyzed via analytical chemistry methods in Sect. 3. Sect. 4 reports the tests of the detectors with cosmic rays, and a direct comparison of the performances in the detection of 2 GeV/$c$ pions during the three periods under investigation. Finally in the Conclusions the results of this work are summarized.

## 2. Detector Configuration

*2.1 Structure of the Detectors*

The configuration of the two aerogel Cherenkov detectors, A1 and A2, is shown schematically in Fig. 1, and their characteristics are listed in Table 1. The concept of this design involves several considerations. The most important one is the maximization of the detection surfaces (PMT's windows) for Cherenkov photons with respect to all other surfaces (diffusive reflector foils and aerogel radiator). Each detector consists essentially of:

- Two separated boxes containing the silica aerogel and the PMT's to facilitate the assembling and the maintenance of the detector;
- Two facing rows of PMT's placed close to the envelope of the particle beam;
- Diffusive reflector material on the inner surface;
- Silica aerogel hydrophobic, with high transparency and suitable refractive index.

The frame of the two boxes is made of aluminium. One box holds the PMT's to the lateral sides of a central empty space (the light diffusion box) while the other accommodates the aerogel radiator. In order to minimize photon losses, all the internal surfaces, with the exception of the PMT's windows and the radiator surface, are covered with reflecting material such that pockets are avoided and the corners are rounded by the reflector, see Fig. 1. In this configuration the PMT's windows are the only internal surfaces detecting the Cherenkov photons and the ratio of PMT's windows to the other surfaces is maximized. Moreover, since the PMT window external annulus of 5 mm radius has low quantum efficiency, it was covered with the reflecting material. To provide light tightness of the whole assembly, O-rings are inserted between the radiator box and the light diffusion box and between each PMT body and its port in the detector structure. Filtered air was flushed inside the detectors to prevent the degradation of PMT due to the presence in the environment of helium released from cryogenic systems. Since 2005 nitrogen has been used in place of the filtered air.

*2.2 Photomultipliers*

Cherenkov light is detected in A1 by twenty-four 5" Burle RCA 8854 [14] PMT's while A2 holds twenty-six 5" Photonis XP 4572B [15]. The Burle Quantacon has good quantum efficiency from the visual to UV range (22.5% at 390 nm), moreover it has a very good pulse height spectrum (high peak to valley ratio) which is very useful for detecting a limited number of PE's. The Photonis XP 4572B is less efficient in the UV region but has better performance in the visual range by ~24%, and especially has higher electron collection efficiency. Comparative tests between Burle and Photonis PMT's have been performed at JLab [16] using a Cherenkov spectrum generated by electrons ($^{106}$Ru source) passing through a small slab ($2.5\times2.5\times1$ cm$^3$) of SP50 aerogel and reflected by Millipore paper. Those measurements show that the PMT quantum efficiency convoluted with the Cherenkov spectrum, the aerogel transparency and the Millipore reflectivity, comprehensively indicated as *QE*, is twice as large in Photonis PMT compared to the Burle one.

In the present design the PMT's are placed much closer to the envelope of the beam particles and positive HV is used. These particulars are implemented in order to maximize the photon collection. The aerogel detectors are operated near the focal plane of HRS where some magnetic field is present. Hence, originally [17] μ-metal shields were installed with the PMT's. However, the magnetic field is low (<0.5 Gauss) and the enhancement of the PE collection was only 5% [18]. At the same time the necessity to move the PMT's outward in the light box and to allocate space for the application of the μ-metal shields resulted in a sizeable reduction of detected PE's. This is because of a larger distance from the light box to the phototube window and a smaller density of PMT's due to this greater distance between their centres. Since the μ-metal installation results in a sizeable reduction in the observed PE's, the detectors are assembled with no magnetic shields in all configurations.

*2.3 Reflectors*

Originally, two 0.45 μm layers of Millipore filter paper GSWP00010 [19] were used as diffusive reflector. Its light reflection is diffusive and follows Lambert's law having 95% diffusivity up to 400 nm, while in the UV it decreases smoothly to 80% at 315 nm. The performance of this material is much better than other reflectors (Teflon, NE560, BaSO$_4$) [20] provided that two or more Millipore papers are put together. One foil, in fact, is partly transparent. The Millipore is fragile, cannot be installed directly, and needed to be backed by common white paper layers using a 3M double sided tape [21] which can be a source of contamination.

During efforts to refurbish the aerogel detectors it was discovered that the Millipore paper had yellowed. Therefore, it was replaced with a new mirror material: the Enhanced Specular Reflector (ESR) by Vikuiti, a 3M company [21]. This material is easy to handle, cut and generally install in narrow and curved regions between the PMT's. Moreover, close to the PMT's diffusion is not an effective process for collecting light while mirror reflection works better. The Millipore was still used with good results in the flat region of the light diffusion box, where the installation of this paper is simple and the use of a diffusive material is more suitable to reflect the light, see Fig. 1.

*2.4 Silica Aerogel*

The HRS select particles having a narrow range in momentum (± 5%). In momentum near 2 GeV/*c* for the identification of pions, kaons and protons, it is possible to design a PID system based on the Cherenkov effect using only two refractive indices. Silica aerogel is the only transparent material that can be employed as radiator in this situation rather than compressed gases. Due to the low self-absorption and Rayleigh diffusion of the Cherenkov light, this material can be used even in a RICH detector [22]. Nowadays several aerogels of good optical quality are commercially available [23]. In our detectors the Matsushita silica aerogel [24] is used mainly because it is hydrophobic and has suitable refractive indices: 1.015 (SP15) in A1, and 1.050 (SP50) in A2. The other aerogel characteristics are reported in Table 2. In the momentum range 1.5-2 GeV/*c*, a firing A1 identifies pions, while a not firing A2 identifies protons. Kaons are identified by a firing A2 and a not firing A1.

The optical quality of aerogel tiles is presented by the transmittance (*T*) curve, well reproduced by the Hunt formula: $T = A \cdot \exp(-Ct/\lambda^4)$, where *T* is the fraction of light transmitted through a *t* thick tile as a function of the wavelength $\lambda$. The Hunt parameters, *A* (transparency) and *C* (clarity, $\mu m^4$/cm units), account respectively for absorbed and Rayleigh scattered light [22]. Samples of high optical quality have *A* values very near to one and low, near to zero, values of *C*.

Fig. 2 reports the transmittance measurements, performed using a PerkinElmer Lambda 3B double-ray spectrophotometer [25], for new SP15 (A1) and SP50 (A2) tiles (dotted areas), inserted in the 2005 restored configuration, and for old tiles of the 2000 original configurations (dashed areas) that were found contaminated during the 2004 visual inspection of the inside of the detectors. The bands in Fig. 2 are obtained by overlapping measurements on different points of the same tile and on different tiles. The higher dotted areas show the superior optical quality of new tiles while the lower transmittance values of dashed areas measure the deterioration suffered by 2000 tiles during the years of operation. The deterioration is greater for the SP15 silica aerogel. Table 2 also lists the average values of the parameters *A* and *C* deduced for the two types of silica aerogel by fitting the measurements of Fig. 2 with the Hunt formula. The quoted uncertainties account for the width of the bands in Fig. 2. On average, old SP15 and SP50 tiles have transparencies 10% and 20% lower, and clarity 300% and 10% higher respectively than new values. The Hunt parameters measured for 2005 tiles are consistent with the quantities found for the 2000 tiles at the time of their installation. Transmittance measurements performed randomly on the old tiles of the inner layers showed no deterioration. On visual inspection these tiles were found to be completely colorless and transparent, confirming that, in both A1 and A2, the degradation was confined to the outer layers. As a trade off between costs, performance, and required time for the refurbishment, only the two outer layers of aerogel have been replaced. Finally the refractive index of old contaminated tiles, measured with the minimum deflection method, was found to be the same. More details on the transmittance and the refractive index measurements are reported in Ref. [26].

Finally it is instructive to estimate in an approximate way the average number of PE's detected in A1 relatively to A2. This quantity depends on four parameters: refractive index of silica aerogel, $n$, the thickness of radiator, $d$, the quantum efficiency of the PMT's convoluted with the Cherenkov light spectrum, the silica aerogel transparency and the Millipore paper reflectivity, $QE$, and the absorption of the Cherenkov light in the radiator. In particular the light absorption is parameterized by $\left[1-\exp(-d/l)\right]$ [27], where $l$ is defined as the effective saturation length and is equal to 3.9-4.0 cm in a wide range of particle momenta. Assuming constant value of radiator index $n$ in usable range of wave length, it is easy to find for $n \sim 1$ and $\beta=1$ particles that the average number of detected PE's is proportional to: $QE(n-1)d\left[1-\exp(-d/l)\right]$. Considering the construction parameters of the detectors, see Tables 1 and 2, and the characteristics of the PMT's, see § 2.2, the number of photons, detected in A2, is expected to be four times higher than in A1. This estimation is in good agreement with the experimental results as will be illustrated in Sec. 4.

## 3. Contamination

In order to determine the causes of the aerogel contamination, several chemical and physical analyses [28] were carried out on several aerogel samples and on the adhesive tape originally suspected of being the main source of pollution in the aerogel. The contaminated aerogel looks yellow because of its increased absorbance in the blue region with respect to the red part of the visible light, see Fig. 2. This effect is evident to the unaided eye when several layers of aerogel are stacked together, but difficult to observe in a single tile.

The reported stability [29,30] of the RICH of the HERMES experiment at DESY (Hamburg), which employs the same aerogel (Matsushita SP30 with $n$ = 1.030), excludes the possibility that the observed degradation could be ascribed to normal aging. Moreover, the same observation is reported in Ref. [31] for the aging induced by thermal variations in 0-100 °C range. The HERMES aerogel optical properties, in fact, have not been deteriorated even after ten years of operation.

As mentioned before the Matsushita aerogel is hydrophobic [24]. The same weight, measured for contaminated and contamination free tiles, suggests that the degradation is not due to absorbed water. Moreover to verify possible differences of hydrophobicity on the tile surface, the measurements of "contact angle" have been performed by means of a goniometer (NRL model 100-00, Ramé-Hart). For new and contaminated aerogel samples, a contact angle of 155° (average value of five measurements) has been observed and therefore no difference has been detected.

Different aerogel tiles (SP15, SP30 and SP50), contaminated and uncontaminated, together with the adhesive tape were investigated by means of Solid-Phase MicroExtraction-Gas Chromatography/Mass Spectrometry (SPME-GC/MS). Small amounts of these aerogels and of the adhesive tape (0.5 - 1.7 grams) were introduced into 40 mL glass vials and heated to ~ 85° C for twenty minutes. Then a SPME fibre (divinylbenzene-carboxen-polydimethylsiloxane) by Supelco [32] was exposed for five minutes into the head space in order to sample the outgassing organic substances. After exposure, the SPME fibre was introduced into the injection port, maintained at ~ 250° C, of a gas chromatograph (Trace GC ultra, Finnigan-Thermo [33]), coupled with an ion trap mass spectrometer (Polaris Q, Finnigan-Thermo). The different molecules of the gas mixture, desorbed from the fibre, are then separated as the sample travels the length of the chromatographic column.

The chromatograms relevant to aerogels SP15, SP30 and SP50 (configuration '04 and '05) showed the predominant presence of styrene and xylene isomers in the tiles, see Fig. 3. Comparison of these chromatograms with that of the degassed fraction of the adhesive tape reveals that only a few of the molecules detected in the aerogel samples are also present in smaller amounts in the adhesive tape. Vice versa the chromatogram of the adhesive tape shows some typical markers such as phenol and methylphenol that are completely absent in the aerogels, see Fig. 3. The histogram in Fig. 4 shows that the styrene/xylene ratio is higher in uncontaminated silica aerogels. These findings could be likely explained by assuming that the air, flushed in the detectors, removes part of styrene used in the aerogel manufacturing.

Finally, as demonstrated in several works [29,30,34], the irradiation of the tiles, due to the beam or radioactive sources, cannot be responsible for the aerogel deterioration.

These results indicate that aging, humidity, irradiation, the tape, and dust are not a significant cause of the observed aerogel contamination. The flushed air with sub-micron particles therein, is the only remaining possible vehicle of the aerogel contamination. In detail the contaminants must consist of sub-micron particles, otherwise they would not pass through the air filter (filters stop molecules larger than 1 micron size) and they would not enter in the silica aerogel matrix (pore size smaller than 100 nm). A further confirmation of this picture is given by the working condition of the HERMES RICH [30] that has kept unaltered its performances along ten years of operation. In that case pure nitrogen, instead of filtered air, has been flushed through the detector.

As illustrated in Fig. 2, the contaminated tiles have a moderate transparency loss in the red part of the spectrum (20%) and strong light absorption in the blue-UV region (larger than 50%). Similar but not identical effects on the aerogel transparency are detected in two other works [31,35]. In those cases the aerogel tiles are contaminated by oil fumes [31] or vapours of solvents (acetone) [35]. These molecules could be some, but not the only causes of our silica aerogel degradation. Anyway it was decided to flush with nitrogen instead of filtered air in the '05 configurations, and to replace the Millipore paper with a plastic material, the Enhanced Specular Reflector [21]. ESR, being plastic, does not produce dust and does not absorb or allow the passage of micro-particles. Yet, it reflects according to the Snell's law with higher reflectivity (98%) in the visible which falls down in the UV region (45% at 390 nm). This lower reflectivity contributes in a negligible way to the PE number reduction of the '05 configuration compared to the '00 original one.

Finally we have intended to detect the elemental carbon through the Thermal Optical Transmittance Technique [36] but physical properties of the silica Aerogel made impossible the measurement using such procedure. A more sophisticated search in order to positively identify the nature of these contaminants is presently not possible. In any case such measurements should be performed using instruments and techniques having very high resolution power (~ 10 nm).

## 4. Experimental Results

*4.1 Cosmic Ray Tests*

Before installing the 2000 detectors in the hadron spectrometer, several tests with cosmic rays were run to quantify the average number of photoelectrons and to perform additional checks on the PMT's and the construction quality [37]. Similar tests were also performed before the '05 installation. The apparatus, used in those tests, is schematically illustrated in Fig. 5. A coincidence between three 15×40×2 cm$^3$ plastic scintillators (S1, S2 and S3), supplies the event trigger while a lead block, 61 cm thick, is used to select relativistic cosmic muons ($\beta \geq 0.994$). The Cherenkov detector was horizontally translated with respect to the trigger system (trigger plus lead block) to test different regions.

A calibrated (pedestal at the ADC channel 0, single PE peak at channel 100) single PMT response is shown in Fig. 6 with the square marker histogram. Since the PE number detected by each PMT ($\mu_i$) is low (≤1), the simplest way to calculate this quantity is by the fraction of counts in pedestal peak *(Po$_i$)* assuming a Poisson distribution for PE's leaving the photocathode:

$$\mu_i = -\ln(Po_i). \qquad i = 1, 2, \ldots. \qquad (1)$$

The whole detector response, achieved summing up all the PMT's calibrated signals, is shown in Fig. 6. For higher numbers of detected PE's (>1) such a distribution can be fitted to the equation:

$$f(q) = const \left\{ \sum_{k=0}^{\infty} \frac{\mu^k e^{-\mu}}{k!} \cdot \frac{1}{\sigma_k \sqrt{2\pi}} \cdot \exp\left(-\frac{(q-kQ_1)^2}{2\sigma_k^2}\right) \right\}, \qquad (2)$$

which is the sum of Gaussians, each representing the contribution of the $k^{th}$ PE, convoluted with the Poisson distribution. In detail, $q$ is the ADC channel, *const* is a constant of normalization, $Q_1$ is the average charge collected at the PMT output for one PE detection and $\sigma_1$ is the corresponding standard deviation, $\mu$ is the mean number of collected photoelectrons. The standard deviations of each Gaussian are linked by: $\sigma_k = \sigma_1 \cdot \sqrt{k}$, while the relative peaks are expressed by the term: $kQ_1$. For $k = 0$ the Gaussian becomes a Dirac's delta function, $\delta(q)$. Eq. (2) is generally used to calculate $\mu$ by fitting the total sum spectra. It is important to notice that use of Eq. (2) requires correct determination of $Q_1$, which is not always done as will be explained in the next section. Finally, we are able to determine the global performance of each detector through the efficiency, ε, calculated according to:

$$\varepsilon = 1 - \exp(-\mu). \qquad (3)$$

Fig. 7 shows the comparisons between the number of PE's ($\mu_i$) recorded with cosmic rays by every PMT in the 2000 and 2005 configurations. The scintillator trigger is positioned in front of the 6$^{th}$ and 18$^{th}$ PMT in A1 (left panel), and the 6$^{th}$ and 20$^{th}$ PMT in A2 (right panel). In Fig. 8 the corresponding total sum amplitudes are shown. The average PE numbers detected by A1 and A2 for the two configurations, obtained by fitting the distributions of Fig. 8 with Eq. 2, are listed in Table 3. All this indicates that the restoration of the threshold detectors is good but not complete, in fact, the reduction of PE number between 2000 and 2005 configurations is ∼ 21% for A1 and ∼ 33% for A2. This result is attributed to the reduced optical qualities of the aerogel tiles which have not been replaced.

Finally it is necessary to notice two effects. The selected cosmic muons ($\beta \geq 0.994$) are above the Cherenkov's threshold both in A1 ($\beta_{th\text{-}SP15} \approx 0.985$) and in A2 ($\beta_{th\text{-}SP50} \approx 0.948$). In particular the number of photons collected in A1 is 7% below with respect to $\beta=1$ particles [38]. This result does not affect the previous conclusions since the same factors have to be included in the analysis of both configurations.

*4.2 In-Beam Tests*

The PE distributions detected by the two detectors for the three configurations analyzed here were acquired using 2 GeV/*c* pions and are reported in Fig. 9. These events are selected in the central acceptance of the two HRS arms imposing several trigger cuts. In particular, among other things, a coincidence between the electron and the hadron recorded in the other HRS arm is required. The particle momentum is set by the kinematic of the running experiment (i.e. the angle between the two spectrometers matched with the beam energy) while the time of flight between the electron and the hadron timing signals identifies the pions. Those selections eliminate a large part of the accidental background, and the events induced by the coincidence of other hadron particles (kaons and protons) having same momenta. More details on the complete data analysis procedures are reported in the Ref. [39]. The continuous curves in Fig. 9 have been obtained by fitting the distributions with Eq. 2. In this distribution each PMT signal of the aerogel detector is calibrated as indicated in Sect. 4.1. The average numbers of PE's, deduced from these fittings, are listed in Table 4.

The comparison between the photoelectron distribution, see Fig. 9, and the PE average numbers, see Table 4, clearly indicates that the best performance of the detectors is obtained in the 2000 original configuration. The 2005 detectors detect a reduced number of PE's, ~ 88% of the original number, both in A1 and in A2. In the 2004 degraded configuration, the PE reduction is more serious both for A1 (~ 60%) and A2 (72%) and consequently the particle identification is appreciably compromised. The comparison of Tables 3 and 4 for A1 also shows that a nearly equal PE number is produced by 2 GeV/$c$ pions and by relativistic muons. This pseudo equality is just fortuitous in A1 as is also evident from A2 results.

Large non-gaussian tails are observed on both sides of the amplitude spectrum, especially for the single PE response of some PMT's (e.g. Burle RCA 8854, Thorn EMI 9351) [40]. Those tails sizeably modify the average value of the pulse height distribution which no longer corresponds to the peak. The identification of the single PE peak is important because it is commonly used to perform the PMT calibration. To account for these tails, Bellamy et al. [41] added an exponential tail to Eq. (2) which was not considered a part of the single PE peak response but a background term. Bellamy's procedure induces an underestimation of the number of detected PE's. The most complete and correct description needs to include those tails in the shape of the PMT response to 1-, 2-, ….$n$-photoelectron and then to convolute this quantity with the Poisson distribution. Those calculations can be performed either using a Monte Carlo [42], or using an analytical procedure [43]. As shown in Ref. [44] when the average number of PE is large (> 4 PE), the results described in latter methods and Eq. (2) (using the correct value of $Q_1$) are very close.

The fitting procedure performed on the PE distributions, using the approach of Dossi et al. [43], has indicated that $\mu$ is ~15% higher for A1 than the values calculated in Table 4. By Eq. (3), A1 and A2 efficiencies do not change their values sizeably because of the relatively high number of detected photoelectrons. Therefore a slight variation of detected PE's in A1 and A2 does not significantly vary the rejection factor of the PID system, which is essentially dominated by electron knock-out.

Finally both cosmic tests and in-beam measurements indicate a reduction for the detected PE's number due to the differences between the 2000 and 2005 configurations. That reduction is sizeable in cosmic tests but is negligible in beam measurements, see Tables 3 and 4. This result is not surprising since it depends on the Cherenkov photon path in the detector. As illustrated in Figs 1 and 5, in the cosmic tests Cherenkov photons suffer more reflections and their path, especially inside the radiator, is longer with respect to the in-beam measurements.

## 6. Conclusions

This paper presents the performance of two aerogel Cherenkov detectors, A1 and A2, in the hadron spectrometer of JLab's Hall A at the time of their first installation (2000), when a reduced PE yield was discovered (2004), and soon after their restoration (2005). In the 2000 original configuration when crossed by 2 GeV/$c$ pions, A1 and A2 detected on average 6.6 and 29.2 PE's respectively. A reduction of these numbers of ~ 62% for A1, and ~ 72% for A2, was observed in the 2004 experiments. The vehicle of the contamination is attributed to the flushed air which introduced sub-micron contaminants absorbed mainly in the outer layers of the radiator. An exact determination of the contaminant's chemical nature by mean of other instruments and or techniques is presently not possible but would be highly appreciated by the community in order to improve the long term operation of these detectors.

In the restored configuration, the outer two aerogel layers were replaced, the Millipore coating was replaced by non-absorbing ESR sheets, and nitrogen instead of air was used as the circulating gas. In this configuration only a ~ 12% of reduction of PE's, for both A1 and A2, was found compared to the original for 2 GeV/$c$ pions. A larger reduction (~ 30%) was found between these yields when measured with cosmic rays. This observation was attributed to the residual effects of the contaminants in the silica aerogel and of the different set-up positions used in cosmic tests with respect to the in-beam measurements. Finally the restoration recovered the exceptional performances of the original detector which is very important for the successful running of several experiments. Based on the previous considerations the new setups will preserve the current characteristics for the long term. This feature is especially important considering the usefulness of this PID system in the future experiments at the upgraded CEBAF accelerator.

# Acknowledgments

The authors wish to acknowledge the constructive discussions with H. Breuer, P. Brindza, and C. Zorn, and J. Segal together with help from the technical staff of INFN-Sanità: S. Colilli, R. Fratoni, F. Giuliani, M. Gricia, M. Lucentini and F. Santavenere, who have contributed in excellent way to this work. Part of this paper is the D. Macchia's Diploma thesis. SM wishes to acknowledge the warm hospitality of the colleagues during his stay at Jefferson Lab. This work has been made possible by grants from Thomas Jefferson National Accelerator Facility which is funded by U.S. Department of Energy under the contract no. DE-AC05-84150, modification no. M175, under which the Southeastern Universities Research Association, Inc. operates the facility, and from the Italian Istituto Nazionale di Fisica Nucleare.

## List of Tables

**Table 1. Main characteristics of the A1 and A2 detectors and the relative thresholds in velocity ($\beta_{th}$) and in momentum for the emission of Cherenkov light by pions, kaons and protons.**

| Detector | Radiator Size (cm$^3$) | Number of PMT's | PMT Model | $\beta_{th}$ | Pions (GeV/$c$) | Kaons (GeV/$c$) | Protons (GeV/$c$) |
|---|---|---|---|---|---|---|---|
| A1 | 170×32×9 | 24 | RCA 8854 | 0.985 | 0.8 | 2.8 | 5.4 |
| A2 | 192×30×5 | 26 | XP 4572B | 0.948 | 0.4 | 1.5 | 2.8 |

**Table 2. Characteristics of the silica aerogel. Hunt parameters are measured in the new (uncontaminated) and in the contaminated tiles. $A$ represents the transparency and $C$ ($\mu m^4$/cm) the clarity of the tiles.**

| Detector | Aerogel Type | Refractive Index, $n$ | Tile Size (mm$^3$) | $A$ contaminated tiles | $C$ ($\mu m^4$/cm) contaminated tiles | $A$ new tiles | $C$ ($\mu m^4$/cm) new tiles |
|---|---|---|---|---|---|---|---|
| A1 | SP15 | 1.015 | 100×100×10 | 0.83±0.04 | 0.0232±0.0032 | 0.95±0.02 | 0.0082±0.0012 |
| A2 | SP50 | 1.055 | 100×100×10 | 0.71±0.05 | 0.0156±0.0027 | 0.90±0.02 | 0.0155±0.0005 |

**Table 3. Average PE number detected by A1 and A2 for two configurations in the relativistic cosmic muons ($\beta \geq 0.994$) tests. The values are obtained by fitting the distributions shown in Fig. 8 with Eq. (2).**

| Configuration | A1 | A2 |
|---|---|---|
| 2000 Original | 6.6±0.9 | 24.7±2.3 |
| 2005 Restored | 5.2±0.8 | 16.4±1.6 |

Table 4. Average PE number detected by A1 and A2 for the three different configurations and for 2 GeV/*c* pions in HRS. The values are obtained by fitting the distributions shown in Fig. 9 with Eq. (2).

| Configuration | A1 | A2 |
|---|---|---|
| 2000 Original | 6.6±0.5 | 29.2±2.1 |
| 2004 Degraded | 2.5±0.3 | 8.2±0.9 |
| 2005 Restored | 5.8±0.4 | 25.8±2.0 |

**List of Figures**

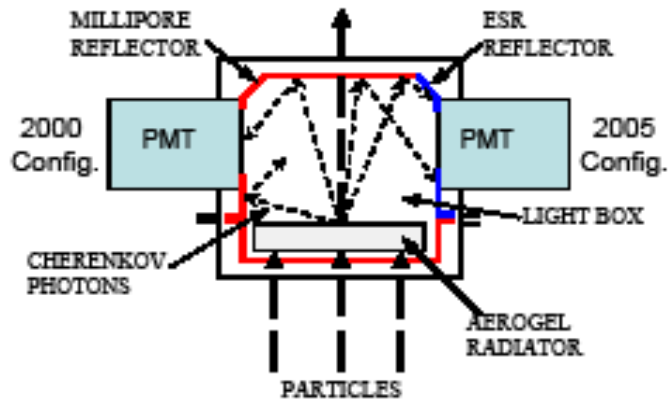

Fig. 1. Schematic configuration (not to scale) of the Cherenkov detector during the beam measurements. The left half of the detector represents the 2000 configuration while the right half draws the 2005 configuration. The difference is the reflector arrangement: Millipore in 2000, Millipore plus ESR in 2005. The paths of the Cherenkov photons are indicated with the dashed lines, read text for more details.

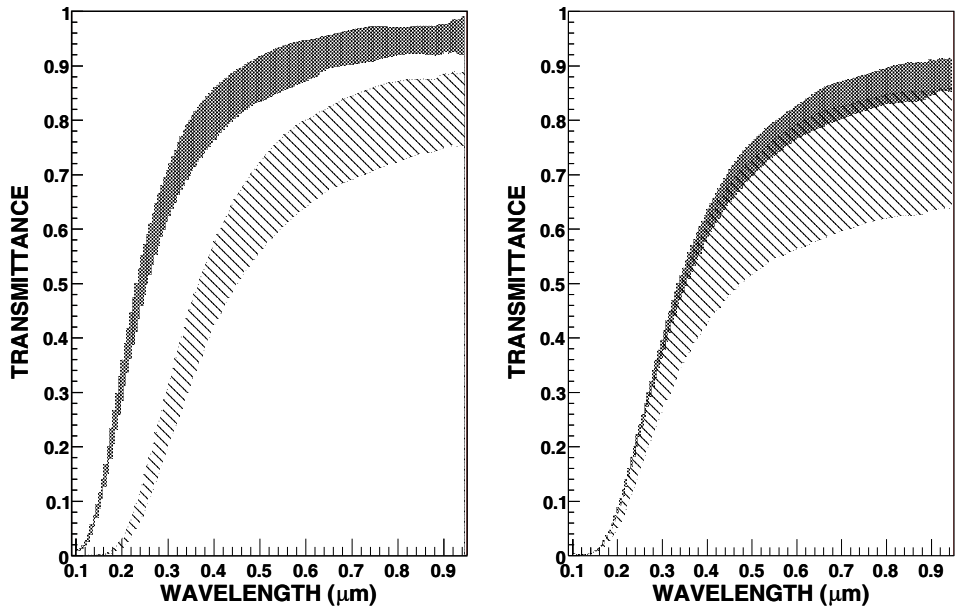

Fig. 2. Transmittance curves of A1-SP15 (left side) and A2-SP50 (right side) aerogel. The bands are obtained by overlapping measurements on different points of the same tile and on different tiles. The dotted areas refer to measurements on new tiles of the 2005 configuration, while the dashed areas represent those of the outer layer of the 2004 configuration.

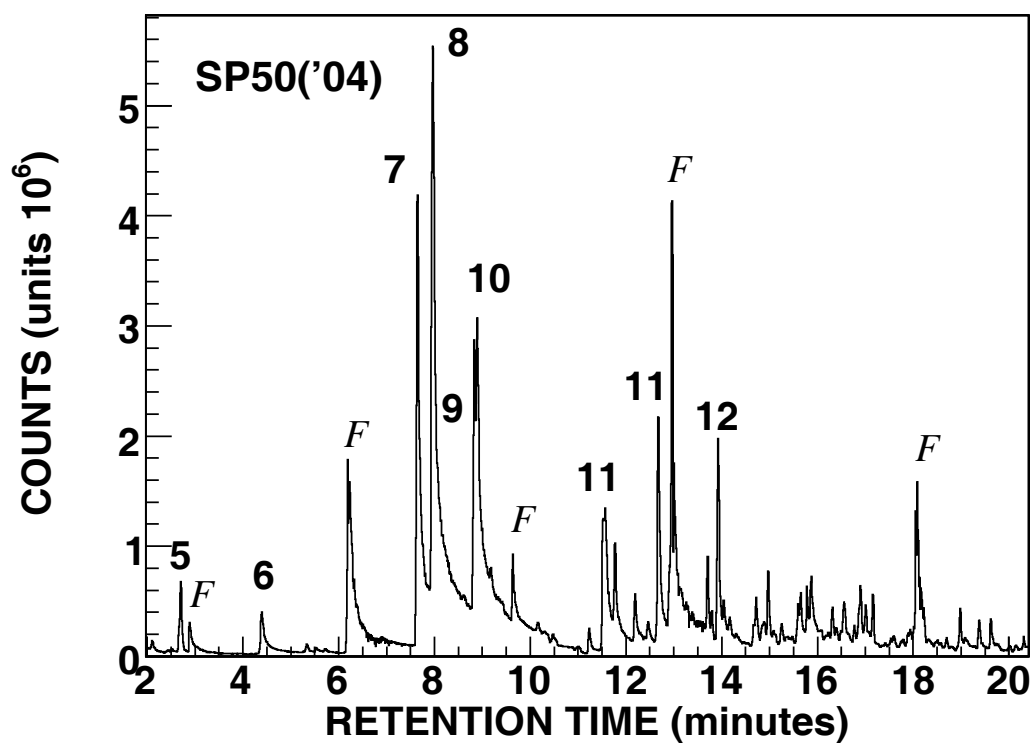

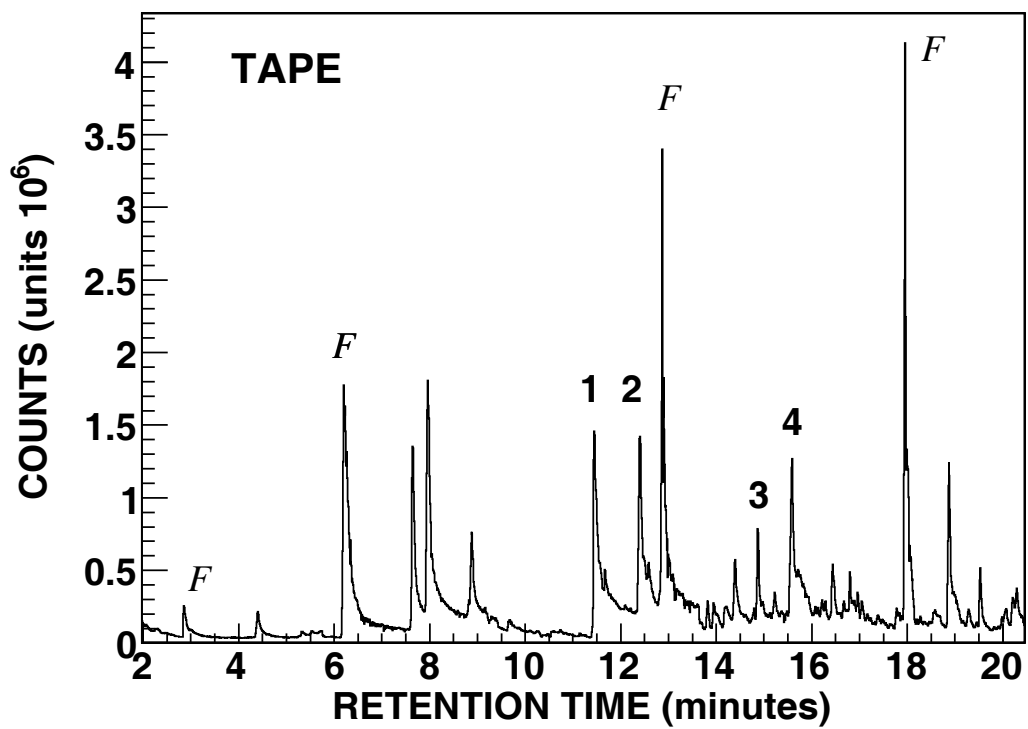

**Fig. 3: Results of SPME-GC/MS analyses performed on adhesive tape (lower panel) and on contaminated SP50 ('04) aerogel tile (upper panel). The main identified compounds are indicated in the chromatograms: benzaldehyde (1); phenol (2); 2-hydroxybenzaldehyde (3); methyl-phenol (4); siloxane (5); toluene (6); *o*-xylene (7); *m*, *p*-xylene (8); styrene (9); ethylbenzene (10); ethyl-methylbenzene (11); monoterpene (12); siloxanes of the SPME fiber (*F*). For the complete list see Ref. [25].**

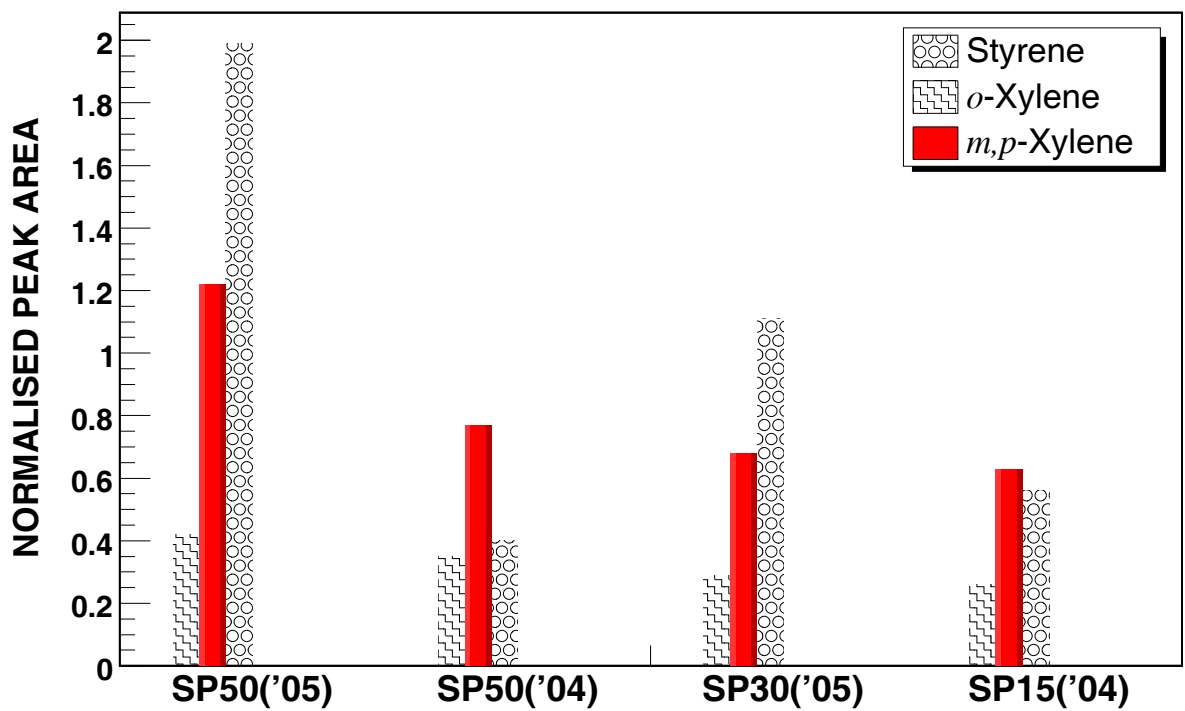

Fig. 4: Peak areas (indicative of mass) of styrene and of xylene isomers (*o*, *m* and *p*) normalised against the peak area of a siloxane typical of the SPME fibre used in this procedure. Normalisation is necessary in order to permit direct comparison between the different aerogel samples. The styrene/xylene ratio is inverted passing from uncontaminated (SP50('05) and SP30('05)) to contaminated (SP50('04) and SP15('04)) aerogels.

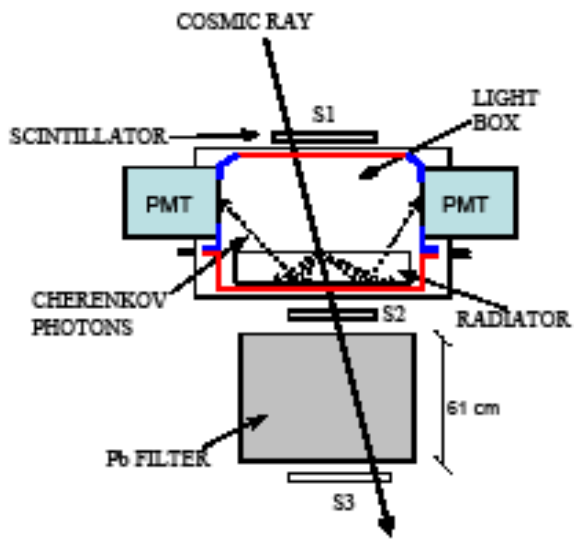

Fig. 5. Schematic view (not to scale) of the set-up used to perform the measurements with cosmic rays. S1, S2 and S3 are the plastic scintillators used for trigger while the Pb filter, 61 cm thick, is used to discriminate the $\beta \geq 0.994$ muons. Before the detection in the PMTs, the Cherenkov photons have to pass through the radiator, travelling longer and suffering more reflections than in the hadron spectrometer. The other details of the drawing are explained in Fig. 1.

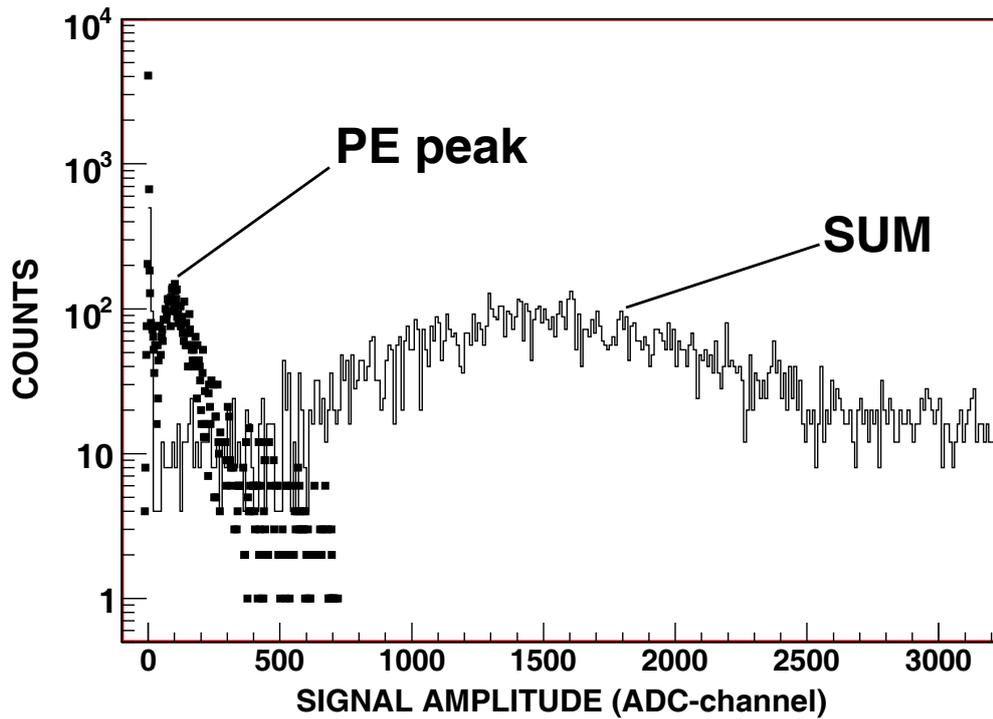

Fig. 6. The sum of amplitudes from all PMTs in A2 is represented by the solid line histogram. The square marker histogram indicates the response of a single PMT. The position of the single PE peak is also indicated because it is used for the calibration of the PMT gain. In our measurements the pedestal is set to ADC-channel 0, and the single PE peak to channel 100.

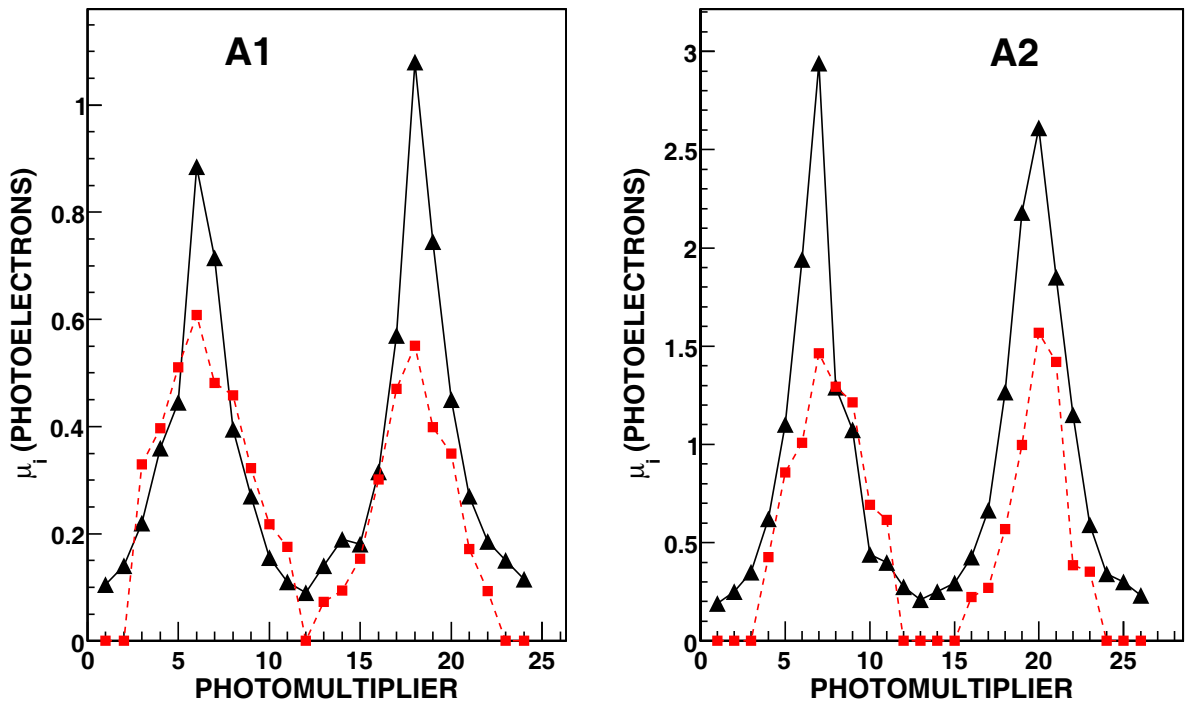

Fig. 7. Number of PE's detected in the cosmic ray test from each PMT of A1 (left panel) and A2 (right panel) in the 2000 original (solid curve and triangle markers) and 2005 restored (dashed curve and square markers) configurations according to the Eq. (1). The curves are only to guide the eye. PMT numbers 1-12 (1-13) are on one side of the A1 (A2) detector, 13-24 (14-26) on the opposite side. The trigger scintillators are positioned in correspondence of PMTs number 6 and 18 for A1, and 7 and 20 for A2.

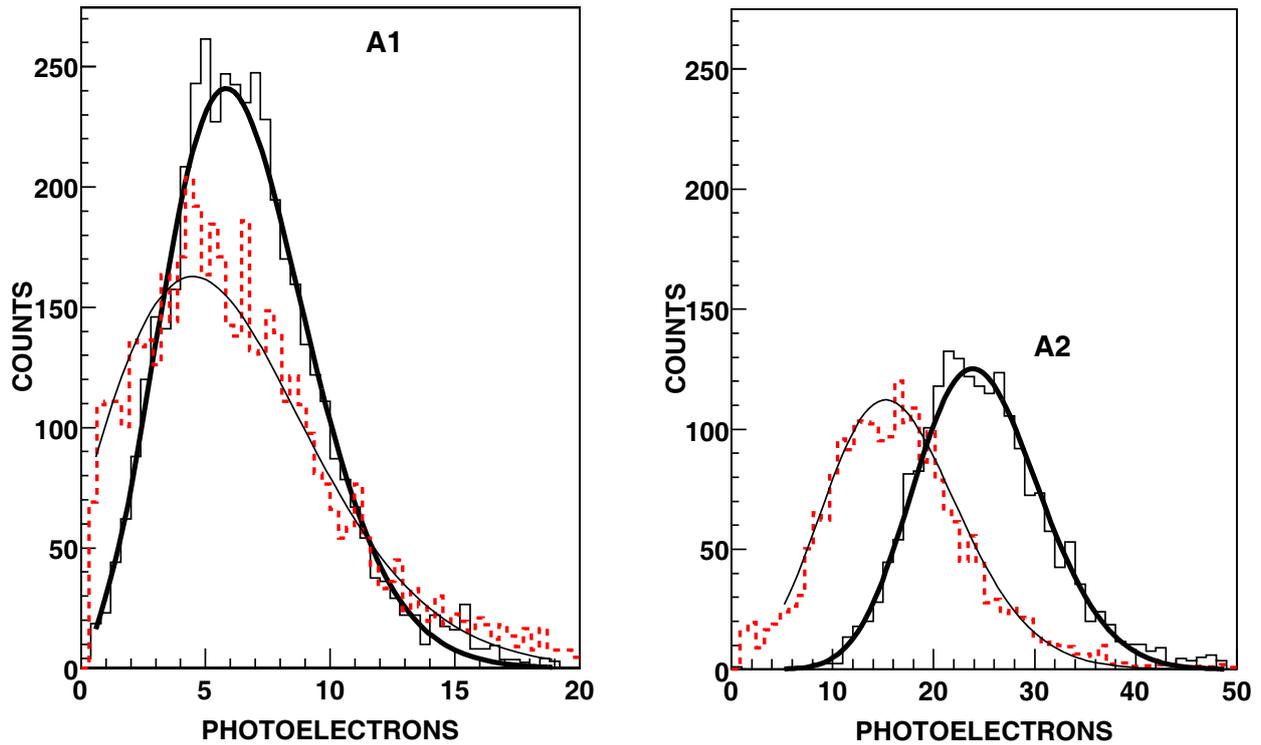

Fig. 8. Distributions of detected PE's by A1 (left panel) and A2 (right panel) in the cosmic ray test. Solid line histograms are for the 2000 original configuration, dashed line histograms for the 2005 restored one. Curves are fit to histograms with Eq. (2). Vertical scales report arbitrary units.

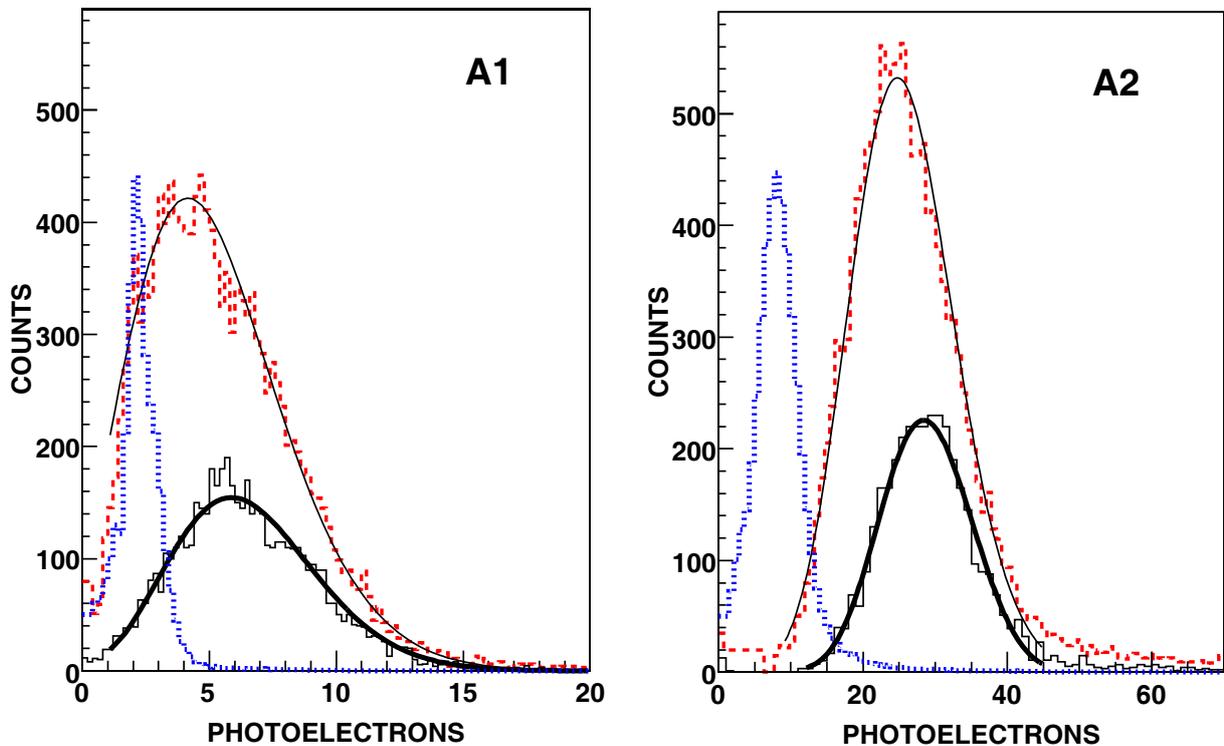

Fig. 9. PE distributions from the sum of all the PMT signals of the A1 (left) and A2 (right) detectors obtained selecting 2 GeV/$c$ pions. Results from the 2000 original, the 2004 degraded, and the 2005 restored configurations are shown with solid, dotted, and dashed line histograms, respectively. Curves are fit to the 2000 and 2005 distributions using Eq. (2) with parameters reported in Table 4. Vertical scales report arbitrary units.